\documentclass[
letter,
twocolumn,
superscriptaddress,
amsmath,
amssymb,
prl,
nofootinbib
showkeys,
10pt,
floatfix,
nobibnotes,
aps,
fltpage
]{revtex4-2}

\usepackage{latexsym}
\usepackage{amsmath}
\usepackage{amssymb}
\usepackage[T1]{fontenc}
\usepackage[open]{bookmark}
\usepackage{hyperref}
\hypersetup{colorlinks=true,allcolors=blue}
\usepackage{upgreek}

\usepackage{orcidlink}

\usepackage{dcolumn}
\usepackage{bm}
\usepackage{xcolor}
\usepackage{hyperref}
\usepackage{cleveref}
\usepackage{float}
\usepackage{svg}
\usepackage{amssymb}
\usepackage{scalerel}
\usepackage{caption}
\usepackage{wasysym}
\usepackage{lipsum}
\usepackage{enumitem}

\providecommand{\pnl}[1]{{\textcolor{black}{(#1)}}}

\newcommand{\red}[1]{\textcolor{black}{#1}}
\newcommand{\green}[1]{\textcolor{black}{#1}}

\makeatletter
\pretocmd\frontmatter@keys@format{\addvspace{20\p@}}{}{}
\makeatother

\usepackage{svg}
\usepackage{cleveref}
\usepackage{siunitx}
\usepackage{caption}
\usepackage{multirow}
\usepackage{makecell}
\usepackage{tabularray}
\usepackage{makecell} 
\usepackage{array} 
\usepackage{graphicx} 
\usepackage{booktabs} 

\begin{document}

\title{Controlling Electron-Beam-Induced Charging in Colloidal Quantum Dots}

\author{Sven Ebel\,\orcidlink{0009-0005-3224-6413}}
\affiliation{POLIMA---Center for Polariton-driven Light--Matter Interactions, University of Southern Denmark, Campusvej 55, DK-5230 Odense M, Denmark}

\author{Alina Myslovska\,\orcidlink{0000-0003-0231-2959}}
\affiliation{Department of Chemistry, Ghent University, Krijgslaan 281-S3, Gent 9000, Belgium}

\author{Stefano~Vezzoli\,\orcidlink{0000-0002-5862-0830}}
\affiliation{
 The Blackett Laboratory, Department of Physics, Imperial College London, London SW7~2BW, United Kingdom
}

\author{Iwan Moreels\,\orcidlink{0000-0003-3998-7618}}
\affiliation{Department of Chemistry, Ghent University, Krijgslaan 281-S3, Gent 9000, Belgium}

\author{Sergii Morozov\,\orcidlink{0000-0002-5415-326X}}
\email{Corresponding author: semo@mci.sdu.dk}
\affiliation{POLIMA---Center for Polariton-driven Light--Matter Interactions, University of Southern Denmark, Campusvej 55, DK-5230 Odense M, Denmark}

\date{\today}

\begin{abstract}
\vspace{0.0cm}
\textbf{Abstract.} 
Colloidal quantum dots (QDs) are attractive nanoscale emitters, yet their response to energetic electron excitation remains poorly understood and often unstable, limiting  cathodoluminescence (CL) spectroscopy, radiation-detection applications, and electron-beam-based device processing.
Here, we investigate \red{the CL response of giant-shell CdSe/CdS QDs and strategies to improve its stability through combined in situ CL and photoluminescence (PL) measurements. 
Spectral and lifetime analysis reveals efficient multiexciton excitation, leading to blueshifted emission and faster dynamics, while charge accumulation contributes to cathodobleaching. 
Using indirect excitation to reach sub-pA effective currents, we demonstrate partial recovery of long-lived neutral-exciton emission and strongly reduced cathodobleaching, although biexciton emission persists even at these low currents.
Furthermore, replacing long insulating ligands with shorter ones strongly suppresses biexciton emission, consistent with improved charge drainage.
Our work presents two complementary methods to control the CL response of QDs under electron-beam excitation,} providing a practical route toward stable CL for spectroscopy and imaging, fast scintillation, and electron-beam-compatible photonic devices.
\vspace{0.3cm}
\end{abstract}

\maketitle

\section{Introduction}

Quantum emitters serve as building blocks of photonic devices, which are often fabricated, inspected, and characterized using electron-beam-based techniques. 
For colloidal quantum dots (QDs), this is a particular challenge because their emission can degrade under electron-beam irradiation, making it necessary to understand the underlying mechanisms and develop strategies to preserve their optical properties~\cite{RodriguezViejo1997,Mahfoud2013,Diroll2025}. 
Cathodoluminescence (CL) microscopy provides a direct way to address this because it combines electron-beam excitation with optical spectroscopy within standard electron-microscopy workflows. 
More broadly, CL is relevant not only for fundamental studies of QDs, but also for their applications such as scintillators, CL imaging probes, and electron-beam-driven optoelectronic devices~\cite{Padilha2013,Woo2013,Fern2015,Turtos2019,Fratelli2024,Guzelturk2024}. 

\red{However, CL from colloidal QDs remains challenging:
compared with photoluminescence (PL), it often exhibits
reduced efficiency, spectral reshaping, fast dynamics,
and cathodobleaching, commonly associated with
multicarrier excitation and beam-induced charging~\cite{Padilha2013,Diroll2025}.
Other possible mechanisms include local heating,
radiolysis and charge-driven atomic motion,
beam-induced surface or ligand modification,
and nonradiative Auger losses~\cite{Mahfoud2013,Padilha2013,Woo2013,Jiang2023,Diroll2025,Oliveira2025}.
Although ballistic displacement of lattice atoms is unlikely at electron-beam energies of 5--30\,keV commonly used
in scanning electron microscopy (SEM)~\cite{Look1974}, the relative contributions of these processes remain difficult to distinguish.
In particular, prompt multicarrier excitation and progressive charge accumulation may coexist and jointly influence the CL response.}

\red{Improving CL stability therefore requires controlling the balance between carrier generation and charge dissipation.
The latter can be enhanced by improving the electrical coupling between QDs and their surroundings, which is strongly influenced by the ligand shell.
While long insulating ligands can hinder charge drainage, shorter ligands, commonly used to improve electronic coupling in QD films~\cite{Rastogi2018,dosSantos2019}, could facilitate charge dissipation under electron-beam excitation.}

\begin{figure*}
\includegraphics[width=0.85\linewidth]{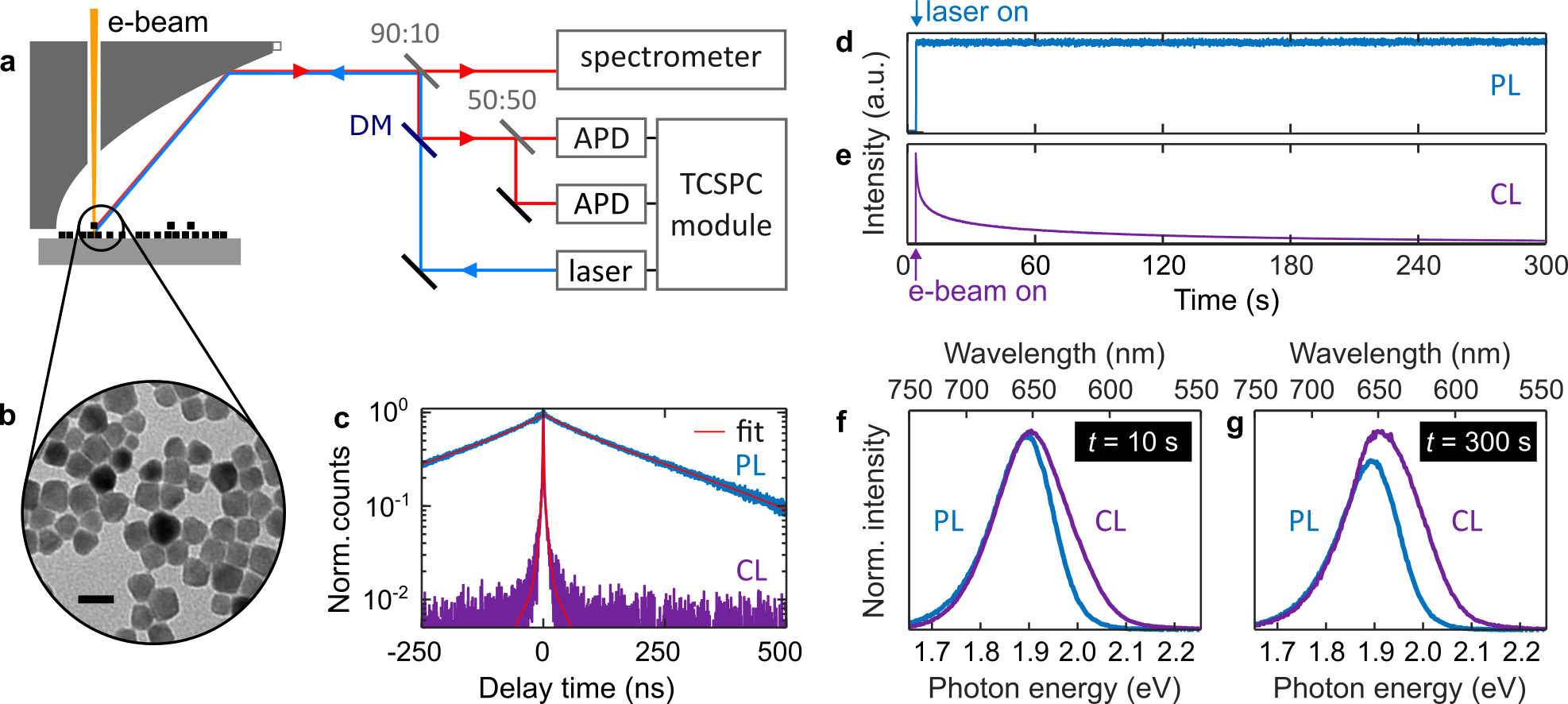}
\caption{\textbf{Experimental setup for in situ CL and PL of QDs.}
\pnl{a}~SEM-based CL setup with a parabolic mirror for light collection. QDs are excited by the electron beam to generate CL, which is analyzed using a spectrometer and a Hanbury Brown--Twiss (HBT) interferometer with two avalanche photodiodes (APDs) connected to a time-correlated single-photon counting (TCSPC) module. A laser is coupled into the same optical path via a dichroic mirror (DM) to excite PL at the focal point of the parabolic mirror.
\pnl{b}~TEM image of giant-shell QDs (scale bar: 25\,nm).
\pnl{c}~Lifetime measurements under electron-beam (CL) and laser (PL) excitation.
\pnl{d-e}~Intensity time traces of QDs under laser and electron beam excitation. Time bin is 10\,ms.
\pnl{f}~Initial CL spectrum at $t=10$\,s showing the appearance of a blueshifted component relative to PL.
\pnl{g}~CL spectrum after prolonged electron-beam exposure of 300\,s, exhibiting a pronounced blueshift compared to PL spectrum. Spectra are normalized at 1.893~eV to facilitate comparison of the high-energy spectral region. \red{The integration time was 1\,s for both PL spectra and the initial CL spectrum at $t=10$\,s, and 10\,s for
the CL spectrum at $t=300$\,s. The CL measurements in Fig.~1(c,e--g) were performed at 39\,pA and 30\,keV.}
}
\label{fig-intro}
\end{figure*}

\red{Related challenges also arise in PL of conventional colloidal QDs, including} reduced photostability under intense laser excitation, blinking due to highly efficient Auger recombination, and the difficulty of isolating exciton and multiexciton contributions~\cite{Klimov2000,Nirmal1996}.
Giant-shell CdSe/CdS QDs have emerged as a photostable alternative with suppressed blinking~\cite{Chen2008,Christodoulou2014}. 
The thick CdS shell suppresses non-radiative Auger recombination and blinking, \red{resulting in} high PL quantum yield, while the quasi-type-II band alignment reduces electron--hole overlap and extends exciton lifetimes to hundreds of nanoseconds. 
Moreover, in combination with interface strain due to the CdSe/CdS lattice mismatch of 4\%,  the wurtzite geometry gives rise to a pronounced biexciton blueshift that can be spectrally isolated even at room temperature, allowing for quantitative assessment of the biexciton contribution by combined spectral and lifetime analysis~\cite{Morozov2023}.

\red{Here, we investigate electron-beam excitation and charging in giant-shell CdSe/CdS QDs by comparing in situ CL and PL spectra and lifetime measurements from the same QD clusters.
Our results indicate that efficient multicarrier excitation contributes to the observed spectral blueshifts and fast emission dynamics, while charge accumulation promotes cathodobleaching.
Using indirect excitation, we find that biexciton emission persists even at sub-pA effective currents, while cathodobleaching can be mitigated and long-lived neutral exciton emission dynamics partially recover.
Furthermore, ligand exchange suppresses biexciton emission,
consistent with enhanced charge drainage.
Together, these findings highlight complementary strategies
to control the CL response of colloidal QDs for electron-beam-based applications, including CL imaging,  scintillators, and electron-beam-compatible photonic devices.}

\section{Results and discussion}

\subsection{Experimental setup and sample}

The in situ CL/PL setup is schematically shown in Fig.~\ref{fig-intro}\pnl{a}.
The sample is positioned at the focal point of a parabolic mirror, where CL is generated by the electron beam and PL is excited by a laser focused through the same mirror~(see details in Methods).
The emitted light is collected by the \red{same} mirror and directed to a spectrometer and an Hanbury Brown--Twiss (HBT) interferometer for spectral and lifetime measurements, respectively.
This configuration enables direct comparison of CL and PL from the same QD clusters under identical chamber conditions.
\red{CL and PL were measured on the same QD clusters using
different excitation geometries: a focused laser for PL
and a focused electron beam for CL, operated either
in stationary or raster-scanning mode (see Methods)}

Colloidal giant-shell CdSe/CdS QDs were synthesized following established protocols~\cite{Christodoulou2014,Morozov2023}. These QDs consist of a CdSe core surrounded by a thick CdS shell, resulting in overall sizes of $\sim15{-}20$\,nm, as shown in the transmission electron microscopy (TEM) image in Fig.~\ref{fig-intro}\pnl{b}.
The nanocrystals were initially capped with long oleate ligands ($\sim1.5{-}2$\,nm), which can electrically isolate individual QDs and hinder charge drainage. 
In subsequent experiments, the native oleate ligands were exchanged for shorter 3-mercaptopropionic acid (MPA) ligands ($\sim0.5$\,nm) using a protocol adapted from previously reported MPA exchange methods for CdSe/CdS nanocrystals~\cite{DiStasio2014}.
Colloidal QD solutions were spin-coated on gold substrates, forming quasi-isolated, micron-sized clusters \green{(SI Fig.~S1)}, with gold providing efficient charge drainage while not contributing to CL background~\cite{Ebel2025}.

\subsection{\red{Evidence of multiexciton emission}}
To interpret CL emission, we \red{compare in situ CL and PL  photon-correlation histograms from the same QD clusters
using the same HBT configuration (Fig.~\ref{fig-intro}(a,c), Methods).
Although the excitation mechanisms differ, photon emission
in both cases originates from electron--hole recombination.
Under pulsed PL excitation, the temporal profiles of the
correlation peaks provide information about the emission
dynamics. In contrast, under continuous electron-beam
excitation, each stochastically arriving electron can generate
multiple electron--hole pairs within a QD or excite several
QDs, producing temporally correlated photons and a single
bunching peak in $g^{(2)}(\tau)$.
While the bunching amplitude depends on the electron current~\cite{Meuret2015,FiedlerWS2,Fiedler2023,Ebel2026}, its temporal decay reflects the emission dynamics
and reproducing the emitter
lifetime~\cite{Meuret2019,SolaGarcia2021,Fiedler2023}.
The histograms in Fig.~\ref{fig-intro}(c) reveal substantially faster dynamics under CL than PL excitation. 
The PL decay is dominated by a long-lived component with a lifetime of
$\sim280$\,ns, while under high-current CL (39\,pA, 30\,keV) the emission exhibits fast components
with characteristic times of 0.75 and 6.9\,ns (Methods and SI Tab.~S1).
These faster CL dynamics are consistent with an increased
contribution from multiexcitonic and charged states~\cite{Padilha2013}, which typically exhibit shorter emission lifetimes~\cite{Galland2012,Morozov2020,Morozov2023},
as further supported by the spectral analysis presented below.}

In contrast to the stable intensity observed under laser excitation [Fig.~\ref{fig-intro}\pnl{d}], the CL intensity of the same QD cluster progressively decreases during continuous electron-beam exposure [Fig.~\ref{fig-intro}\pnl{e}], demonstrating cathodobleaching. 
This intensity degradation is not caused by local beam-induced heating, as similar behavior persists under continuous cooling at 84\,K \green{(SI Fig.~S2)}.
The CL spectrum evolves during electron beam irradiation: immediately after the beam is turned on, the CL spectrum largely follows the PL peak, but already shows a blueshifted component [Fig.~\ref{fig-intro}\pnl{f}]. 
Prolonged irradiation leads to a decrease in CL intensity accompanied by a pronounced blueshift and spectral reshaping [Fig.~\ref{fig-intro}\pnl{g}].
\red{In contrast, short electron-beam exposure causes reversible PL quenching, with the emission intensity recovering when the electron beam is switched off \green{(SI Fig.~S3)}.
This recovery strongly supports the presence of beam-induced charging, which contributes to the observed changes in the intensity response.
The observations in Fig.~\ref{fig-intro}\pnl{e-g} are consistent with multiexciton generation and progressive charge accumulation.
The former contributes to the spectral blueshift through
state filling and repulsive exciton--exciton interactions~\cite{Padilha2013}, while the latter can alter recombination pathways, enhancing nonradiative losses and reducing the emission intensity~\cite{Diroll2025}.}

\begin{figure}[t]
\includegraphics[width=0.99\linewidth]{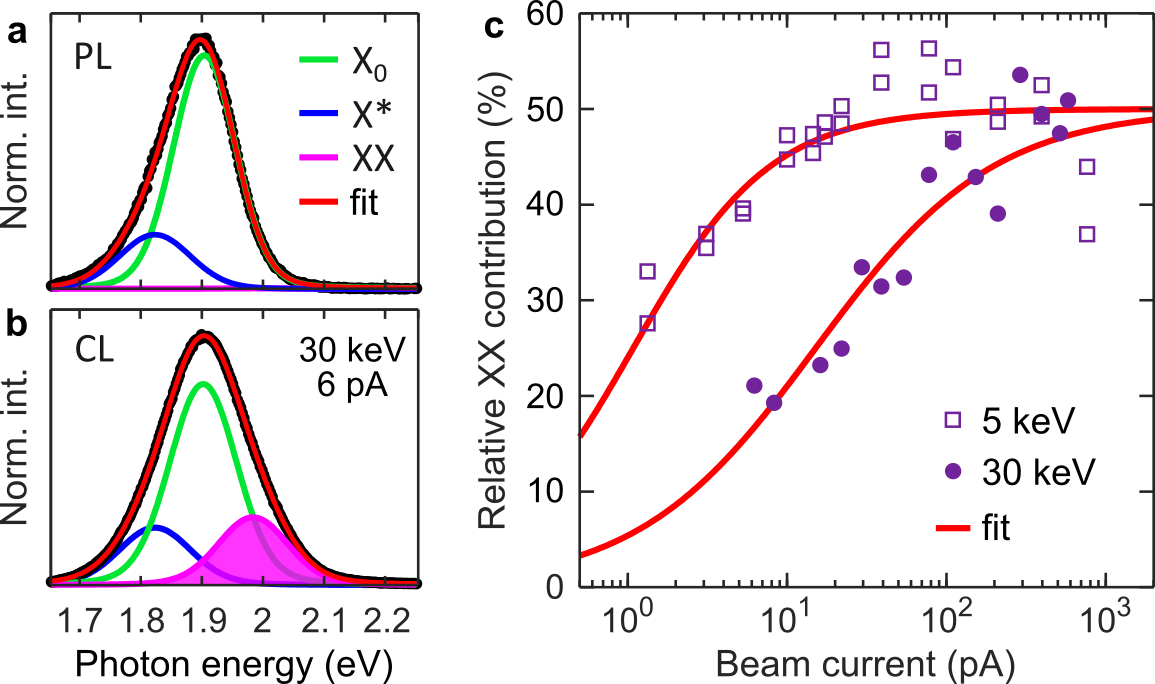}
\caption{\textbf{Biexciton contribution in CL.}
\pnl{a}~PL spectrum decomposed into neutral exciton (X$_0$), charged exciton (X$^*$), and biexciton (XX) contributions with corresponding cumulative fit.
\pnl{b}~CL spectrum decomposed into the same components, showing enhanced biexciton emission (magenta).
\pnl{c}~Relative biexciton contribution as a function of beam current for 5\,keV and 30\,keV excitation.
}
\label{fig-XX}
\end{figure}

To explore how electron-beam excitation drives QDs into the multiexciton regime, we track the current-dependent evolution of the CL spectra using spectral decomposition fitting (Methods).
We analyze CL spectra \red{in the energy domain using a three-component Voigt fit, guided by PL spectral decomposition and supported by three-exponential fits of the decay curves (see SI Tab.~S1, Fig.~S4, and Tab.~S2).}
This enables robust assignment of the neutral-exciton (X$_0$), charged-exciton (X$^*$), and biexciton (XX) contributions.
The PL spectrum in Fig.~\ref{fig-XX}\pnl{a} is well described by a dominant X$_0$ contribution at \red{1.905\,eV (651\,nm)} and a weaker X$^*$ contribution at \red{1.810\,eV (685\,nm)}, where X$^*$ includes trions as well as higher charged excitons.
No detectable XX contribution is observed under these PL conditions because the low-power pulsed laser excitation (500\,kHz, 10\,nW before entering the SEM) is distributed over many QDs within the cluster, and the 2\,$\upmu$s pulse separation is much longer than the \red{$\sim280$\,ns} neutral-exciton lifetime, minimizing XX generation.
Fig.~\ref{fig-XX}\pnl{b} presents CL \red{spectrum and spectral decomposition fit} under low-current electron-beam excitation (6\,pA, 30\,keV), where the X$_0$ and X$^*$ contributions are \red{ constrained within specified ranges around the values} extracted from the PL spectral fit. 
In contrast to PL, an additional high-energy spectral component appears at \red{1.965 eV (631\,nm),} which we assign to XX emission. 
Its \red{60~meV ($20$\,nm)} blueshift relative to X$_0$, matching the XX shift reported for single QDs PL from the same synthesis batch in our previous work~\cite{Morozov2023}, confirms its XX origin. 
At higher beam currents, a weak additional high-energy emission feature becomes visible \green{(SI Fig.~S5)}.

Next, we evaluate how the biexciton contribution in CL depends on the electron-beam parameters, namely beam current and energy [Fig.~\ref{fig-XX}\pnl{c}].
The relative contribution of each component is calculated from the integrated area of its fitted \red{Voigt component}, normalized by the total fitted peak area.
Notably, biexciton emission is evident even under the weakest electron-beam excitation conditions in Fig.~\ref{fig-XX}\pnl{b}, where its relative contribution is \red{21\%}. 
The biexciton fraction increases with beam current and exhibits saturation behavior in Fig.~\ref{fig-XX}\pnl{c}. 
\red{We use a Hill-type function to parameterize this trend, with the saturation level constrained to a maximum of 50\% (Methods). This limit is motivated by an ideal radiative biexciton cascade, in which each XX photon is followed by an X photon, so the XX contribution cannot exceed half of the total cascade emission.
To characterize the high-current regime independently of this constraint, we calculate the mean XX spectral contribution from the measurements within the highest 30\% of beam currents, obtaining $47.9\pm5.5$\% at 5\,keV and $48.1\pm5.5$\% at 30\,keV, where the uncertainties represent standard deviations.
A representative high-current CL spectrum with an XX contribution of approximately 50\% is shown in SI Fig.~S4, with the corresponding fitting parameters summarized in Tab.~S2.
We note that the fitted XX spectral contribution does not directly represent the biexciton generation probability, as it also depends on radiative and nonradiative processes and the populations of other emissive states.
The extracted spectral contributions are also subject to
model-dependent systematic uncertainty associated with
the assumed line shapes and fitting constraints
(see SI Fig.~S4 and Tab.~S2).}

The saturation is reached at significantly lower beam currents for 5\,keV excitation. 
At lower accelerating voltages, electrons have a reduced penetration depth and deposit their energy within a smaller volume, leading to stronger interaction with the sample~\cite{Ebel2025}. 
This confinement increases the local carrier density and is compatible with a more efficient biexciton generation, whereas at higher voltages the energy is distributed over a larger volume, reducing biexciton excitation.
The pronounced biexciton contribution is \red{consistent with
efficient multicarrier excitation, potentially involving
carrier multiplication via impact ionization}
~\cite{Padilha2013}.
Note that, to minimize effects of cumulative charging and cathodobleaching in spectral measurements, CL spectra were recorded from fresh QD clusters in raster-scanning mode, with selected beam currents repeated on two clusters to assess cluster-to-cluster variability and reproducibility (Methods). 

\subsection{Ultra-low-current regime}
\begin{figure} \includegraphics[width=0.99\linewidth]{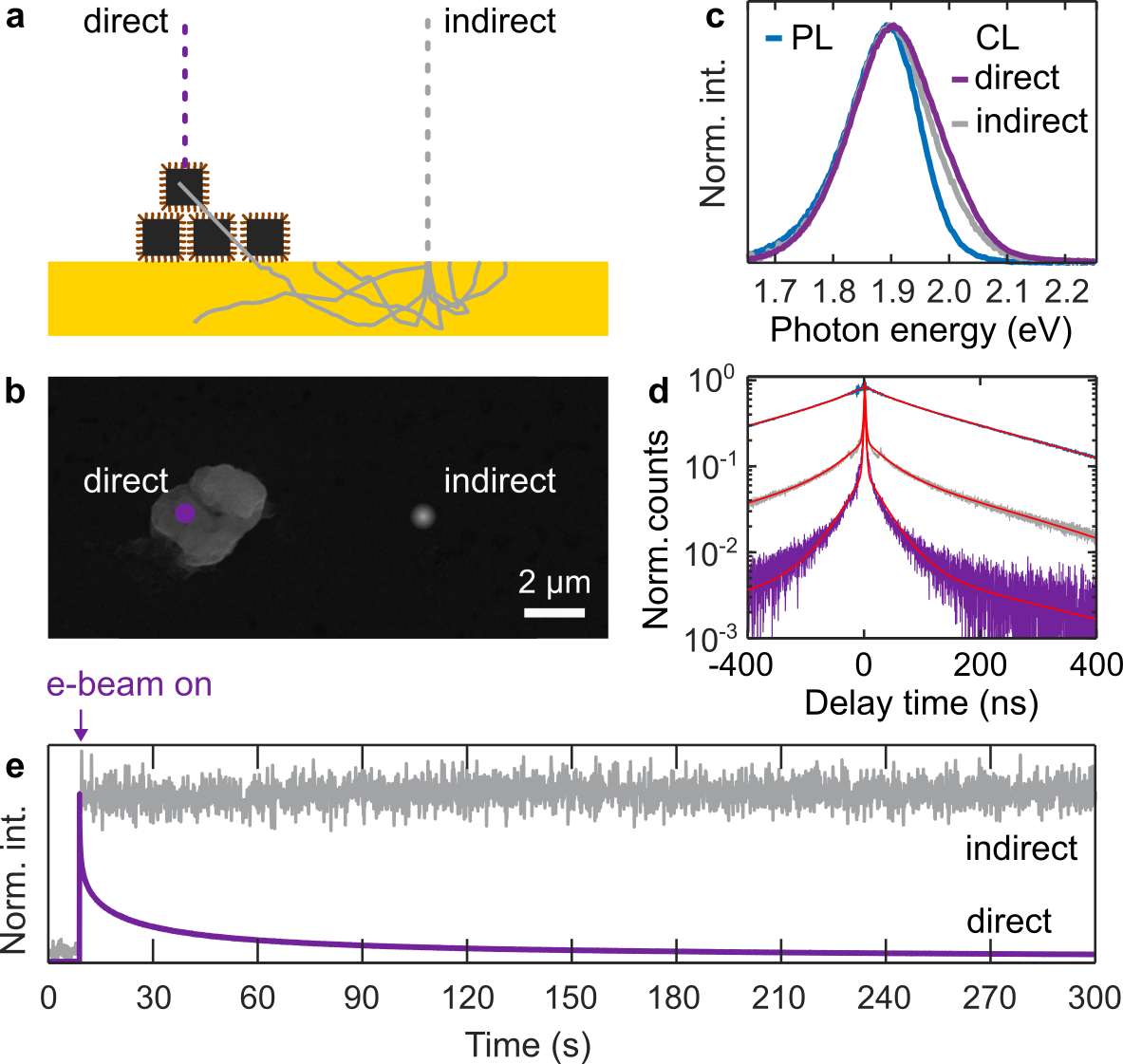} \caption{\textbf{CL response at ultra-low currents.} 
\pnl{a}~Schematic of direct (purple) and indirect (gray) excitation of QDs. 
\pnl{b}~SEM image of a QD cluster indicating beam positions for direct (purple) and indirect excitation 5\,$\upmu$m away from the cluster (gray). 
\pnl{c}~Emission spectra under direct and indirect (5\,$\upmu$m) excitation compared to PL. 
\pnl{d}~Decay histograms in PL (blue), direct (purple) and indirect (grey, at 5\,$\upmu$m) CL. 
\red{The CL measurements in panels (c) and (d) were performed
at the same beam parameters of 6\,pA and 30\,keV.}
\pnl{e}~Normalized intensity time trace under direct excitation at 6\,pA showing bleaching, and under sub-pA indirect excitation at 9\,$\upmu$m showing stable emission.
} 
\label{fig-ind} 
\end{figure}

We next reduced the electron beam current to test whether cathodobleaching can be suppressed and to examine which emitting states are excited in QDs at low currents.
Even at the lowest instrument-accessible current of 6\,pA at 30\,keV, the CL intensity degrades during continuous exposure [purple in Fig.~\ref{fig-ind}\pnl{e}].
To overcome the instrument limitation on lowest current, we employ substrate-assisted CL, or indirect excitation~\cite{Iyer2023,Ebel2026}, where the QDs are excited not by the primary electron beam but by substrate-generated electrons, reducing the  effective excitation current down to the sub-pA regime. 
Fig.~\ref{fig-ind}\pnl{a} schematically shows the direct and indirect excitation mechanism. 
In indirect excitation, the primary electrons impinge on the substrate, generating secondary and backscattered electrons (SEs and BSEs), which can propagate toward the QD cluster. 
Among these, BSEs dominate due to their higher energies and longer propagation lengths, enabling excitation over micrometer distances~\cite{Ebel2026}. 
\red{We note that substrate-generated SEs and BSEs can contribute to both direct and indirect excitation, although their contribution in direct excitation is expected to be significantly smaller than that of primary electrons.}
In Fig.~\ref{fig-ind}\pnl{b}, SEM imaging indicates the beam positions used for direct excitation (purple) on the QD cluster and indirect excitation (gray) on the nearby gold substrate.
This allows us to probe the same QD cluster under direct excitation at 6\,pA and under indirect excitation 5\,$\upmu$m away, where the effective current is reduced by approximately two orders of magnitude~\cite{Ebel2026}.

 \begin{figure*}
\includegraphics[width=0.99\linewidth]{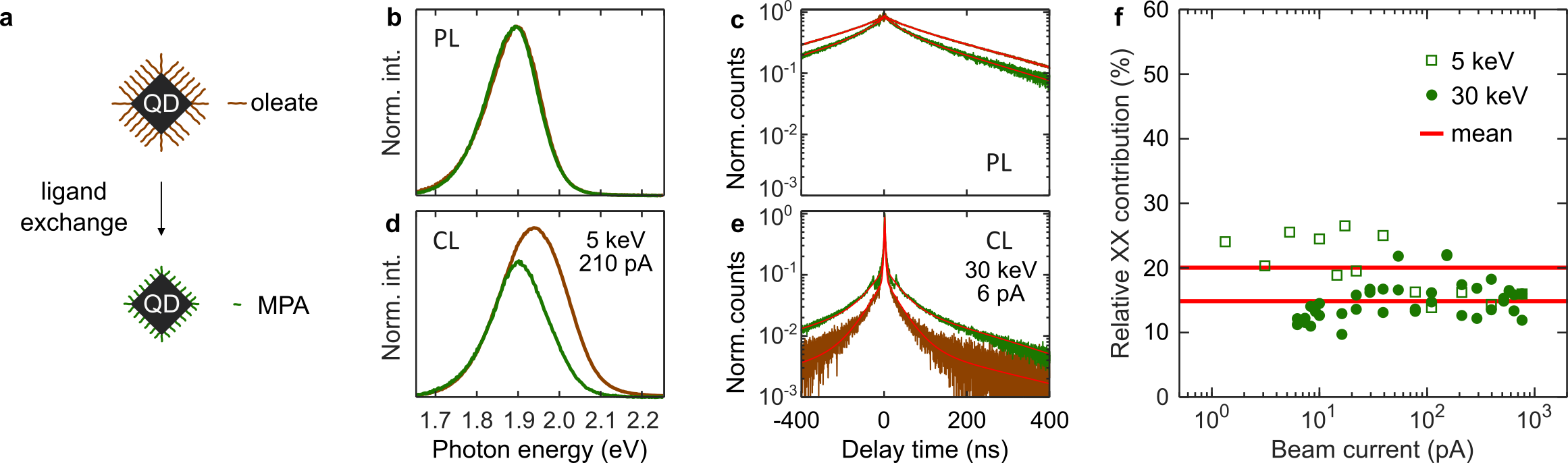}
\caption{\textbf{Effect of ligand exchange on biexciton generation.}
\pnl{a}~Schematic of ligand exchange from long oleate ligands (brown) to short MPA ligands (green).
\pnl{b,c}~PL spectra and decay histograms before and after ligand exchange.
\pnl{d}~High-current CL spectra, demonstrating suppression of the biexciton emission after replacing long oleate ligands with shorter MPA ligands.
Both spectra are normalized at 1.879~eV to highlight the contribution of biexciton emission.
\pnl{e}~Low-current CL decay histograms before and after ligand exchange, together with three-exponential fits (red).
\pnl{f}~Relative biexciton contribution after ligand exchange over a wide range of electron-beam currents for 5\,keV and 30\,keV. 
}
\label{fig-LE}
\end{figure*}

To place the excitation conditions on an absolute scale, we estimate the mean number of incident electrons arriving within one exciton lifetime $\tau$ as $N_e = I\cdot\tau/e$, where $e$ is the elementary charge~\cite{Meuret2015}.
This yields almost $10$ electrons per neutral-exciton lifetime at 6\,pA under direct excitation, compared with only $0.1$ electrons under indirect excitation.
Thus, indirect excitation corresponds to a regime well below one electron per lifetime on average, whereas direct excitation at a few pA already places the QDs in a regime of repeated excitation within a single lifetime.
However, the CL spectra in Fig.~\ref{fig-ind}\pnl{c} remain nearly identical: even under indirect excitation, at only $\sim0.1$ \red{electrons} per neutral-exciton lifetime on average, a pronounced biexciton contribution persists, although weaker than under direct excitation.
Increasing the indirect excitation distance to 9\,$\upmu$m further reduces the effective current, yet the biexciton emission does not disappear~\green{(SI Tab.~S1 and Fig.~S6)}.
This highlights the fundamentally different excitation mechanism of CL, where a single energetic electron can generate multiple electron--hole pairs via carrier multiplication processes, leading to multiexciton excitation even in the sub-single-electron-per-lifetime excitation regime~\cite{Padilha2013,Meuret2015}.

The effect of reduced excitation current on the emission dynamics is more clearly seen in the time-correlated measurements shown in Fig.~\ref{fig-ind}\pnl{d}.
The decay curves were analyzed using the three-exponential fitting procedure, excluding small side peaks from detector afterglow as described in Methods~\cite{Fiedler2023,becker2005advanced}, and the extracted lifetimes and relative contributions are summarized in \green{SI Tab.~S1}.
The characteristic lifetimes remain nearly unchanged across PL, direct CL, and indirect CL, with a fast component of $\sim1.9$\,ns \red{consistent with multiexcitonic recombination, potentially involving biexcitons and
higher-order multiexcitons}, an intermediate component of $\sim40$--50\,ns related to charged excitons, and a long neutral-exciton component of $\sim260$\,ns.
The primary effect is therefore not a change in the recombination lifetimes, but a redistribution of their relative contributions, further supporting our interpretation that the CL dynamics is governed by charge generation and accumulation rather than by irreversible surface degradation.
Under direct CL excitation (purple), the long component accounts for \red{$\sim33\%$} of the total signal, whereas the fast and intermediate components contribute \red{$\sim21\%$} and \red{$\sim46\%$}, respectively.
In contrast, under indirect CL excitation (gray) at 9\,$\upmu$m away from the QD cluster, the contribution of the long-lived component increases to $\sim85\%$, approaching the PL value of $\sim95\%$ (blue). 
Nevertheless, even under low-intensity indirect excitation, the initial decay still \red{exhibits a fast component}, confirming that bright colloidal nanocrystals with strong \red{multiexciton} emission are promising systems for ultrafast timing in scintillators~\cite{Lecoq2017}.
Finally, Fig.~\ref{fig-ind}\pnl{e} shows the temporal evolution of the CL intensity. Under direct excitation at 6\,pA, the emission gradually bleaches. Indirect excitation at 5\,$\upmu$m still shows some cathodobleaching \green{(SI Fig.~S6)}, whereas at 9\,$\upmu$m [gray in Fig.~\ref{fig-ind}\pnl{e}] the CL intensity stabilizes over time.
Overall, these results show that ultra-low-current excitation can mitigate charge accumulation and cathodobleaching, while biexciton generation remains present even in the sub-single-electron-per-lifetime regime.

\subsection{Charge drainage}

To examine the role of charge accumulation in multiexciton formation, we control charge dissipation through ligand exchange (Fig.~\ref{fig-LE}). 
We improved the electrical coupling between the QDs and the substrate by replacing long oleate ligands ($\sim 1.5{-}2$\,nm) with shorter MPA ligands ($\sim 0.5$\,nm), as schematically shown in Fig.~\ref{fig-LE}\pnl{a} (Methods).
The PL spectra in Fig.~\ref{fig-LE}\pnl{b} before and after ligand exchange remain nearly unchanged, indicating that the intrinsic emissive properties of the QDs are preserved.
The corresponding PL decay histograms in Fig.~\ref{fig-LE}\pnl{c} show only a modest redistribution of the relative contributions, possibly due to changes in the QD surface environment after ligand exchange, while the long-lived neutral-exciton component remains dominant \green{(SI Tab.~S1)}.
In contrast, the CL response changes markedly after ligand exchange. 
Under severe charging conditions at low accelerating voltage and high beam current, the CL spectra in Fig.~\ref{fig-LE}\pnl{d} show a pronounced suppression of the biexciton emission for the MPA-treated sample, already visible directly from the reduced high-energy shoulder. The effect is also reflected in the CL decay dynamics at the low current in Fig.~\ref{fig-LE}\pnl{e}, where ligand exchange increases the relative weight of the longer-lived components, consistent with reduced biexciton contribution. 
The stronger spectral suppression of the biexciton shoulder in Fig.~\ref{fig-LE}\pnl{d} suggests that lifetime analysis at high current would reveal an even larger difference.
However, because the $g^{(2)}(0)$ bunching amplitude scales as $1/I$~\cite{Meuret2015}, high-current measurements have a low signal-to-noise ratio, especially for the long lifetime components, which are therefore lost in the uncorrelated background (Methods).
\red{Both spectral and temporal measurements in Fig.~\ref{fig-LE}\pnl{d-e} show suppressed
multicarrier emission following ligand exchange, regardless
of excitation parameters, supporting improved charge drainage.}

We then repeat the spectral decomposition analysis introduced in Fig.~\ref{fig-XX} for the MPA-capped QDs. 
Following ligand exchange, the biexciton contribution is lower and already saturated within the investigated beam-current range [Fig.~\ref{fig-LE}\pnl{f}].
We therefore quantify the biexciton level using the mean and standard deviation, obtaining \red{$20\pm 5$\%}  at 5\,keV and \red{$15 \pm 3$\%} at 30\,keV, as indicated by the red horizontal lines in Fig.~\ref{fig-LE}\pnl{f}.
\red{The substantial reduction of the biexciton fraction following
ligand exchange supports the role of charge accumulation
in enhancing multiexciton emission, alongside prompt
multicarrier excitation.}
By facilitating charge dissipation, the shorter ligands reduce the likelihood of multiple-carrier occupation within individual QDs, \red{although changes in surface passivation and trap densities may also contribute to the observed suppression of biexciton emission.}

\red{Beyond fundamental CL studies, controlling charge accumulation
and emission stability in QDs is relevant to scintillators,
CL imaging probes, electron-beam-driven optoelectronic devices,
LEDs, and bioimaging~\cite{Padilha2013,Woo2013,Fern2015,
TuanNguyen2023,Singh2023,Golovynska2026}.
For scintillation in particular, fast emission must be balanced
with high light yield and stability under energetic
excitation~\cite{Turtos2019,Fratelli2024,Guzelturk2024,
Katznelson2026}.
Extending our findings from QD clusters on gold to individual
QDs and thicker scintillator films will be important for
assessing their potential in practical applications.}

\section{Conclusion}

We have characterized the optical response of colloidal giant-shell CdSe/CdS QDs under electron-beam excitation.
By comparing and decomposing the spectral and lifetime features of CL and PL from the same QD ensembles, we found that CL contains a pronounced biexciton contribution, even when QDs are excited at the lowest currents available in standard electron-beam systems.
The biexciton contribution reflects the high excitation density produced by electron-beam excitation, \red{potentially involving the generation of multiple electron--hole pairs per QD through carrier multiplication via impact ionization}~\cite{Padilha2013}.
Cathodobleaching is likewise linked to charge accumulation over time.

We further tested this hypothesis in the ultra-low-current regime, below 0.1\,pA, using indirect excitation of the QDs.
Under these conditions, cathodobleaching is strongly reduced and \red{the long-lived neutral-exciton component} partially recovers, approaching the PL response.
To investigate charge accumulation further, we \red{performed a simple ligand exchange, demonstrating suppressed biexciton emission,
consistent with enhanced charge drainage. } 
Together, these results identify efficient multiexciton generation and charge accumulation under electron-beam excitation as key factors governing the CL response of colloidal QDs.
This provides practical routes for CL optimization through control of excitation conditions and surface chemistry, paving the way for QD operation in CL spectroscopy, imaging, scintillation, and electron-beam-compatible photonic devices.

\section*{Methods}

\textbf{Lifetime measurements.} PL and CL emission dynamics were characterized using second-order correlation measurements acquired in an HBT configuration (Fig.~\ref{fig-intro}a). The detected photons were split by a 50:50 beamsplitter and recorded by two avalanche photodiodes (APDs, SPCM-AQRH-14-TR, Excelitas Technologies) connected to a time-correlated single-photon counting (TCSPC) module (Time Tagger 20; Swabian Instruments), yielding photon arrival-time-difference histograms, $g^{(2)}(\tau)$. \red{All PL and CL photon-correlation histograms were acquired with an integration time of 300\,s.}

For PL measurements, the QDs were excited with a pulsed laser (PicoQuant LHD-P-C-405; 405\,nm, 500\,kHz, 10\,nW before entering the SEM), resulting in a train of correlation peaks separated by the laser repetition period of 2\,$\upmu$s. 
The PL decay dynamics were quantified by fitting the central correlation peak together with the two neighboring peaks at $\pm2$\,$\upmu$s using a shared three-exponential model,
$
I(t)=\sum_i A_i e^{-|t-t_0|/\tau_i},
$
replicated at each laser repetition period. This procedure accounts for the periodic excitation while constraining all three correlation peaks to share the same decay parameters.

For CL measurements, the QDs were excited by a continuous electron beam. In this case, the $g^{(2)}(\tau)$ histogram exhibits a central bunching peak around zero delay, which contains information about the emission dynamics under stochastic excitation~\cite{Meuret2019,SolaGarcia2021,FiedlerWS2,Fiedler2023}. Before fitting, the CL histograms were normalized by the average far-delay level at $|\tau|\geq2$\,$\upmu$s, such that the long-delay baseline corresponds to $g^{(2)}(\tau)=1$. The CL decay dynamics were then quantified by fitting the central bunching feature with a three-exponential model,
$
g^{(2)}(\tau)=1+\sum_i A_i e^{-|\,\tau-t_0\,|/\tau_i},
$
where $t_0$ accounts for a small experimental offset of the zero-delay position. 
To avoid spurious fitting contributions from instrumental artifacts (APD afterglow~\cite{Fiedler2023,becker2005advanced}), data points in a narrow window around $|\tau|\approx23$--44\,ns were excluded from the decay fit.

For visualization only in Fig.~\ref{fig-intro}\pnl{c}, Fig.~\ref{fig-ind}\pnl{d}, and Fig.~\ref{fig-LE}\pnl{c,e} correlation histograms were additionally processed by subtracting the far-delay level evaluated around $\tau=\pm1$\,$\upmu$s and normalizing the resulting signal to its maximum value, facilitating direct comparison of the temporal profiles. All fitting was performed on the non-background-subtracted data. The relative contribution of each decay component was calculated from the fitted parameters as
$
C_i={A_i\tau_i}/{\sum_j A_j\tau_j},
$
which represents the relative contribution of each exponential component to the total integrated \red{fitted
correlation} signal.
\red{The robustness of the three-exponential fitting procedure,
including sensitivity to the excluded delay windows,
is demonstrated in SI \green{Fig.~S7} using representative CL
decay histograms from Fig.~\ref{fig-LE}\pnl{e}.
Note that fitted contributions describe the relative weights of the correlation components and do not directly represent the populations or emission yields of individual excitonic states.}

\textbf{Spectral measurements.}
Point-mode measurements were performed by parking the electron beam at a fixed position on a QD cluster, as in the intensity traces in Fig.~\ref{fig-intro}\pnl{e} and Fig.~\ref{fig-ind}\pnl{e}. 
Because this mode leads to stronger cumulative charging, CL spectra were instead acquired in scanning mode over $20\times20$\,$\upmu\mathrm{m}^2$ windows, so that the beam continuously rastered rather than dwelled on the cluster. 
With $N\times N = 768\times768$ pixels and a dwell time of $t_\mathrm{d}=1$\,$\upmu$s per pixel, the frame time was $t_\mathrm{frame}=N^2 t_\mathrm{d}\approx0.59$\,s, meaning that any given point was exposed for only 1\,$\upmu$s per pass and revisited once every $\sim0.59$\,s. For $I=6$\,pA, the corresponding areal dose per frame, $D=It_\mathrm{frame}/(eA)$ with $A=400$\,$\upmu\mathrm{m}^2$, is $D\approx5.5\times10^4~\mathrm{e}^-/\upmu\mathrm{m}^2$.
A typical CL spectrum was acquired over 1\,s.

\red{\textbf{Spectral fitting.} Emission spectra were converted from wavelength to photon energy using $E=hc/\lambda$, with the spectral intensity corrected by the corresponding Jacobian, $I_E=I_\lambda\lambda^2/(hc)$, where $E$ is the photon energy, $\lambda$ is the wavelength, $h$ is Planck's constant, $c$ is the speed of light, and $I_E$ and $I_\lambda$ are the spectral intensities per unit energy and wavelength, respectively~\cite{Mooney2013}.
The spectra were analyzed using a three-component Voigt model to separate contributions from neutral excitons (X$^0$), charged excitons (X$^*$), and biexcitons (XX). The X$^*$ component includes trions as well as higher charged excitonic complexes.
We selected Voigt profiles as they provide the best fit to the experimental spectra by accounting for both homogeneous and inhomogeneous broadening~\cite{vanDerBok2020} (see SI Fig.~S4 and Tab.~S2).
The relative emission contributions were obtained by integrating
the fitted Voigt components in energy over the spectral range
of 550--750\,nm.
The X$^0$, X$^*$, and XX peak positions were constrained to $1.905\pm0.003$, $1.810\pm0.013$, and $1.965\pm0.003$\,eV, respectively, with Gaussian-sigma-equivalent Voigt width parameters restricted to $53\pm6$, $60\pm10$, and $62\pm6$\,meV.
}

\textbf{Biexciton saturation analysis.}
The dependence of relative biexciton contribution   on beam current $\mathrm{XX}(I)$ was parameterized using a Hill-type saturation function, $\mathrm{XX}(I) = \mathrm{XX}_{\mathrm{sat}}\, I / (I + I_{1/2})$, \red{where $\mathrm{XX}_{\mathrm{sat}}$ is the saturation level, $I$ is the beam current, and $I_{1/2}$ is the half-saturation current.
The saturation level was constrained to $\mathrm{XX}_{\mathrm{sat}}\leq 0.5$, based on the theoretical limit of an ideal radiative XX--X cascade.
}

\textbf{Ligand exchange.}
Ligand exchange was performed on spin-coated QD films by replacing long oleate ligands with shorter 3-mercaptopropionic acid (MPA) ligands.
The ligand-exchange solution consisted of methanol with 1\,vol\% MPA and 0.5\,vol\% ammonium hydroxide, prepared fresh before use.
The QD-coated substrates were immersed in the exchange solution for 10 minutes (dip method), followed by rinsing in clean methanol (2$\times$20\,s) and drying under nitrogen flow.

\noindent
\textbf{Acknowledgments:}
We thank Torgom~Yezekyan for assistance in sample preparation. 
TEM measurements were performed at the UGent TEM Core Facility.
We also thank N.~Asger~Mortensen for fruitful discussions, and Yonas~Lebsir for assistance with the instrument setup. 

\noindent
\textbf{Competing interests:}
The authors declare no competing financial or non-financial interests.

\noindent
\textbf{Funding Declaration:} The Center for Polariton-driven Light-Matter Interactions (POLIMA) is sponsored by the Danish National Research Foundation (Grant No.~DNRF165).
The work presented here is supported by the Carlsberg Foundation (Grant CF24-2081).

\noindent
\textbf{Author contributions:} The idea of the project was conceived by S.~E and S.~M. 
QDs were synthesized by A.~M. and I.~M.
Sample preparation and CL spectroscopy measurements were handled by S.~E and S.~M. 
The project was supervised by S.~M.
All authors contributed to analyzing the data and writing the manuscript. 
All authors have accepted responsibility for the entire content of this manuscript and approved its submission.

\bibliography{bibliography}
\bibliographystyle{unsrt}

\end{document}